\documentclass{article}
\newcommand \m {M$_\odot$}

\newcommand\kms{km~s$^{-1}$}

\def\gcc{ {\rm g \, cm^{-3} } }

\twocolumn

\begin{document}

\noindent{\bf Resource Letter OTS-1: \\
Observations and Theory of Supernovae \\
\medskip
J. Craig Wheeler \\
Department of Astronomy, University of Texas at Austin,\\
1 University Station, \#C1400,\\ Austin, TX, 78712-0259}

\begin{abstract}
This Resource Letter provides a guide to the literature on the
observations of supernovae and the theory of their explosion
mechanisms.  Journal articles and books are cited for the
following topics: observations of the spectra, spectropolarimetry,
and light curves of supernovae of various types, theory of
thermonuclear explosions, core collapse,
and radioactive decay, applications to cosmology, and possible
connections to gamma-ray bursts.

\end{abstract}

\section*{I. INTRODUCTION}

    Supernovae represent the catastrophic explosions
that mark the end of the life of some stars.  The ejected mass
is of order 1 to 10 solar masses (\m) with bulk velocities ranging from a
few thousand to a few tens of thousands of \kms, with a small mass
fraction moving even more rapidly in some cases.  The typical
kinetic energy of the explosion is about $10^{51}$ ergs, an
energy comparable to the binding energy of the central core
of a star and hence evidence that the star is substantially, if
not totally, disrupted.  The explosions eject heavy elements
that seed the gas of the interstellar medium.  Some supernovae
may produce compact remnants such as neutron stars or black holes.
Supernovae also produce extended remnants as their ejecta
drive shock waves into the interstellar medium.  This energy
can produce turbulence in the interstellar medium, help to
form new stars, and  expel matter entirely out of host galaxies
in some circumstances.  Because of their great brightness,
supernovae can be used to determine distances and
constrain the age, shape and dynamics of the Universe and the
evolution of matter within it.  Recent evidence suggests that
cosmic gamma-ray bursts may be linked to supernova-like explosions.

\subsection*{ I.1 History}

Historical records show that seven or eight supernovae have
exploded over the last 2000 years in that portion of the Galaxy
not obscured by dust, about 15 \% of the area of the Galactic plane
(Ref. 14).
The supernova of 1006 was the brightest  ever recorded, brighter
than Venus and perhaps as bright as a  quarter Moon.  The
supernova of 1054 produced the expanding cloud of  gas that
modern astronomers identify as the Crab Nebula.  The  supernova
of 1572 was observed by Tycho Brahe and one in 1604 by
Johannes Kepler.  Of these events, only SN 1054 produced an
observable compact remnant, the Crab pulsar.  Although there is
still some  controversy, the events of 1006, 1572, and 1604 are
generally  thought not to have left any compact remnants, but
to have resulted in complete disruption of the star.

Given the present size and rate of expansion of the remnant,
we deduce that the explosion that gave rise to the strong radio
source Cassiopeia A occurred in about 1679 (Ref. 70 and citations therein).
No bright optical outburst was seen.   This explosion may have been under
luminous, but the degree is  difficult to ascertain since there is a great
deal of patchy dust  extinction in this direction.  There are reports that
Cas
A was  seen faintly by John Flamsteed, the first Astronomer Royal of
England, 
but there are questions concerning whether or not that sighting was
in the same position as the remnant observed today.  Supernovae tend
to be brighter if they explode within large red giant envelopes.
The suspicion is that the star that exploded in about 1680 may have
ejected a major portion of its envelope before exploding or that the
star was otherwise relatively small and compact.  That condition, in turn,
may have prevented Cas A from reaching the peak brightness characteristic
of most supernovae.  Supernova 1987A, the best studied supernova of
all time, had this property of being intrinsically dimmer than usual
due to a relatively compact progenitor star.

We observe the blast waves of the expanding remnants that have been
produced by supernovae (Ref. 81).  The youngest of
these correspond to the historical supernovae, but we also observe the
effects of several hundred older supernova remnants in our Galaxy
and in nearby galaxies that have exploded over the last million years
or so.   Nearly 1000 of the perhaps $10^9$ neutron stars in our Galaxy
have been observed as radio pulsars and binary X-ray sources.  Of the
perhaps $10^8$ stellar mass black holes that have been created over the
history of the Galaxy by core collapse with or without attendant
supernova explosions, about two dozen stellar mass black holes
have been detected in binary star systems.
 
\subsection*{ I.2 The Modern Era}
    
All supernovae observed since the advent of the telescope and
other modern astronomical instrumentation have been in other
galaxies (Ref. 59).   Supernovae occur roughly once
per hundred years for  spiral galaxies like the Milky Way (
Refs. 43 and 8),
but astronomers now systematically  observe a large number of
galaxies to great distances in order  to discover supernovae.
The current rate of discovery is  several hundred per year.
The current record redshift is  z = 1.7, corresponding to a
lookback time of order 10 billion years (Ref. 114).

Categories of supernovae are traditionally defined by the spectrum
that reveals the composition, velocity and other properties of the
ejected matter.  Complementary information is obtained from the
light curve, the pattern of rapid brightening and slower dimming
displayed by each event.  The two basic spectral categories are
called Type I for those that reveal no spectral evidence for
hydrogen and Type II for those that do (Refs. 101 and 102).   The
progenitors
of Type I supernovae are  thought to have lost their outer hydrogen
envelopes
by a combination  of transfer to a binary companion or stellar winds.

Events in a principal sub-category of Type I supernovae are
called Type Ia (Refs. 68, 18, 17 and 50).   Type Ia show characteristic
elements in the spectrum, especially a strong line of once-ionized silicon
(Refs. 196 and 141), but also magnesium,
sulfur, and calcium near maximum light when there is a well-defined
photosphere in the outer layers (Ref. 65) and
iron in later phases when the ejecta are optically thin
(Ref. 90).  The light curves for Type Ia supernovae are
also characteristic.  There is an initial rise to a peak that takes
about two weeks and then a long slower period of gradual decay over
time scales of months that is very similar for all these events.
The data recorded by Tycho and Kepler suggest that they both witnessed
Type Ia supernovae.  For decades, all Type Ia supernovae were thought to
be virtually identical (although see Ref. 111),
but more recent careful observations have revealed quantifiable
variations among them (Ref. 107).  Events that
are intrinsically dimmer near maximum light decline more
rapidly from maximum and show characteristic spectral features,
especially blended lines of ionized titanium (Refs. 93 and 76).

Type Ia supernovae appear in all kinds of galaxies -- elliptical, spiral,
and irregular.  Type Ia tend to avoid the arms of spiral galaxies
(Refs. 97, 60 and 98).   Since
the spiral arms are sites of new star formation, Type Ia
are deduced to explode in older, longer-lived stars.   This is
consistent with their being the only type to explode in elliptical
galaxies that lack recent star formation.  There is evidence that
the subluminous variety of Type Ia are especially old and hence
constitute the population of Type Ia in elliptical galaxies and the
older events in spiral galaxies where stars have a range in ages
(Ref. 87).  Whether this requires separate evolutionary
channels or variations on a single theme is not known.

Spectropolarimetry provides a tool to measure the geometry of
supernovae.  Purely spherically symmetric configurations will
show no net linear polarization.  Such observations of Type Ia
supernovae show that the majority are only slightly polarized, if at all
(Ref. 88).  They are thus substantially spherical explosions.
Some exceptions are known.  SN 1999by, a subluminous Type Ia, and hence
presumably old, showed finite polarization with a rather well-defined
axis of orientation (Ref. 88).  This may be a clue to the
explosion mechanism, for instance suggesting that the subluminous
Type Ia are especially rapidly rotating.

Near the peak of their light output, Type II supernovae show approximately
``solar" abundances in their ejected material, including a normal
complement of hydrogen (Refs. 101 and 102).
An important sub-class, the Type IIn, show narrow emission lines
and other evidence of the collision of the supernova ejecta with
low velocity circumstellar matter shed by the progenitor star before
explosion (Ref. 115).

In the later optically-thin phases, Type II still show hydrogen
features, but also strong emission lines of oxygen, magnesium
and calcium, characteristic of the evolved inner portions of
a massive star (Refs. 17 and 50).
The optical light curve of a typical Type II supernova shows a rise
to peak brightness in a week or two and then a period of a month or
two when the light output is nearly constant (Ref. 50).
After this time, the optical luminosity drops
suddenly over a few weeks and then more slowly with a time
scale of months.  This pattern of light emission with time
is consistent with an explosion in the core of a star with a
massive, extended red giant envelope (Refs. 150, 154 and 56).  The Type IIn
tend to show a long, slower decline with the light output coming
predominantly from the circumstellar interaction (Ref. 145).

Type II supernovae have never been observed in elliptical galaxies.
Type II supernovae occur occasionally in irregular galaxies, but mostly in
spiral galaxies and then within the confines of the spiral arms
(Refs. 43 and 8).  The interpretation
is that the stars that make Type II supernovae are  born and die near
their birth sites and hence in a short time.
Since they are short-lived, the stars that make Type II
supernovae must be massive.  For theoretical reasons based on
the nature of the late stages of stellar evolution, Type II
supernovae are thought to arise in stars with mass in excess
of about 8 solar masses, but observations and associated
statistics are not sufficiently precise to confirm this number.

Two other varieties of hydrogen-deficient supernovae are labeled
Type Ib and Type Ic (Refs. 68, 131, 110, 18, 17 and 50).   The two types are
probably closely related. Type Ib show evidence for helium in the spectrum
near
maximum light, whereas helium has never been identified in a
Type Ia.  Type Ic show little or no such evidence for helium.
Near maximum, Type Ic display iron lines that qualitatively
resemble Type Ia at certain phases, but which can
be distinguished from Type Ia quantitatively, especially
with time series of spectra.  Type Ib and Ic show little or
no evidence for the strong line of once-ionized
silicon that is a major characteristic of the spectra of Type Ia.
Both Type Ib and Ic show evidence for oxygen, magnesium and
calcium at later, optically-thin, phases in a manner similar to
Type II (Refs. 79, 18, 17 and 50).   Type Ia show essentially
only iron at later times, another factor emphasizing their difference
from Types Ib and Ic.  The composition  revealed by Types Ib and Ic
is similar to that expected in the core of a  massive star that has
been stripped of its hydrogen.  In the case of Type Ic, most of the
helium is gone as well.  The light curves of Type Ib and Ic are somewhat
similar to those of Type Ia, but are generally dimmer at maximum light
(Refs. 131 and 50).

Unlike Type Ia, but like Type II, Types Ib and Ic only seem to explode
in the arms of spiral galaxies (Ref. 126).  Type IIn,
Ib and Ic  are often strong radio sources whereas no radio emission
has yet been detected from a Type Ia (Refs. 44 and 45).  Radio emission is
thought to be created when the supernova shock collides with a surrounding
medium, so this radio emission may indicate strong mass loss from Type Ib/c,
again consistent with an origin in a massive stellar core.
Although circumstantial, the preponderance of the evidence
from the location of the events, their composition and the
evidence for appreciable circumstellar matter, is that Types Ib
and Ic are associated with massive stars.
    
The dividing lines of the supernova classification taxonomy are blurred
by events that share some properties of Type I and some of Type II.
A bright supernova observed in 1993, SN 1993J, gave yet more
clues to the diversity of processes that lead to exploding stars
(Refs. 95, 51 and citations therein).  SN 1993J
revealed hydrogen in its spectrum, so this event was a variety of Type II.
As the explosion proceeded, however, the strength of the hydrogen features
diminished and strong evidence for helium emerged.  In this phase, SN
1993J resembled a Type Ib.  Apparently this star had most, but not all, of
its hydrogen envelope removed, probably in a binary mass-transfer process.
There is yet no direct observational proof for binary companions in Type Ib
or Type Ic or the transition events like SN 1993J, but this seems likely.
Strong winds from massive stars could play a role in stripping the
hydrogen and helium from Type Ib and Type Ic and the relative importance
of winds versus binary mass transfer has not been resolved.

Spectropolarimetry has revealed that all supernovae of Type II or Ib/c
or the transition events like SN 1993J are polarized and hence
asymmetric to an appreciable extent (Ref. 129).
Furthermore, the polarization tends to get larger at later phases when
greater depths are observed (Refs. 129 and 94).  Type Ib/c also tend to be
more highly polarized  and hence asymmetric than Type II.  This evidence
suggests that it is the inner portions of the explosion, and hence the
mechanism itself, that is strongly asymmetric.  In many cases (SN 1993J is
a conspicuous exception) the geometry seems to follow a single axis
as if the explosion were bipolar (Ref. 129).

Estimates of the rate of formation of neutron stars in the Galaxy
(Ref. 96) are similar to estimates of the rate of
formation of Type II and Type Ib/c supernovae.  This does not prove that
these supernovae produce neutron stars, but the notion that the two
processes are directly related is a nearly universal working hypothesis.

\subsection*{ I.3 Explosion Mechanisms}
    
Type II and Type Ib/c supernovae thus represent the explosion of
massive stars.  These stars have presumably evolved from the main
sequence to become red giants and have had a series of
nuclear burning stages  that produce ever heavier elements
in the core.  The masses of the stars that give rise to these
events is estimated to be between about ten and perhaps thirty to
fifty solar masses.  Below the lower limit, stars make white dwarfs
that do not explode as single stars, but may explode as Type Ia
supernovae if they occur in binary systems with appropriate mass exchange.
Above the very uncertain upper limit, stars probably collapse to make
black holes with no associated supernova explosion.  The statistics of
the rate of explosion of Type II and Type Ib/c events (Ref. 66)
 is roughly consistent with this estimate.
Stars with initial mass in excess of about 12 solar masses are thought
to form iron cores that collapse to form neutron stars (Refs.
190 and 56).   There is thus a strong
suspicion that Type II and Type Ib/c supernovae make neutron stars as
compact remnants of the explosion and that the gravitational
energy liberated in forming the neutron star is the driving
force of the explosion.  There is a direct relation between
supernovae and neutron stars in some cases, as illustrated
in the Crab nebula with its pulsar.  The long-sought compact remnant
produced by Cas A was discovered in the dramatic first image taken by
the Chandra X-Ray Observatory (Ref. 103).  This object is
quite dim, a factor of about $10^4$ dimmer than the Crab pulsar and
its nature is still uncertain.  At this writing, suspicions of a
rotation period that would confirm a neutron star have not
been verified, and the possibility that it is a feebly accreting
black hole has been entertained.  The spectral types of both the Crab
and Cas A supernovae are ambiguous.  There is yet no evidence for a
compact remnant from SN~1987A.  Other, older supernova remnants
also show direct or indirect evidence for pulsars, but again
there is no evidence as to the spectral type of the event.
No supernova of known spectral type has produced an observed, identified
compact remnant 

At the lower end of the mass range suspected to contribute to Type
II and Type Ib/c supernovae, the evolution may be slightly different.
Computer calculations show that for stars with original mass between about
8 and 12 \m\ carbon burns to produce neon and magnesium, but the oxygen
does not get hot enough to burn (Ref. 190).
The core, composed of oxygen, neon, and magnesium, contracts,
to form a white dwarf.  The neon and magnesium are susceptible to
electron captures at the central densities attained by these
stellar cores.  This process reduces the electron density that is
responsible for the degeneracy pressure that supports the core.
The result is that the core collapses before any of the elements
in the core begin thermonuclear burning.  During the collapse,
the remaining nuclear fuels -- oxygen, neon, and magnesium -- are
converted to iron (Refs. 159 and 138).
The net result is a collapsing iron core, just as for the more massive
stars where the iron core forms before the collapse ensues,
but quantitative differences could affect the explosive outcome.

The iron core that forms in the center of a massive star is endothermic;
it can only absorb energy either by breaking down into lighter elements or
forging heavier elements.  The absorption of energy reduces the pressure
support and guarantees that the iron core will become unstable to
collapse shortly after it forms, in less than a day (Ref. 167).
In practice, the processes that destabilize the core are electron capture
and  photodisintegration of the iron into helium nuclei and free
neutrons (Ref. 56).

In the collapse of an iron core, the protons in the iron-like
nuclei undergo electron capture to become neutrons.  Each such
reaction creates a neutrino. The generation and transport of neutrinos
produced in this and related processes are a critical ingredient
in the process of core collapse (Ref. 147).  Inside a
certain neutrino-trapping radius where the density exceeds about
$10^{11}$ $\gcc$, the matter collapses nearly adiabatically and hence
homologously so that the velocity is proportional to the radius and
the density profile is self-similar.
(Refs. 140, 153, 5 and citations therein). When the collapse
reaches the density of nuclear matter, about $10^{14}$ $\gcc$, the strong
nuclear force develops a repulsive component.  The result of the increased
pressure is that the collapse halts.  The inner homologously collapsing core
settles into hydrostatic equilibrium and a shock forms at the
boundary and propagates out into the still infalling matter.
The binding energy of the neutron star, about $10^{53}$ ergs,
must be radiated away in neutrinos.  This is 100 times more than
is necessary to blow off the outer layers containing
calcium, oxygen, carbon, and helium, and any outer envelope of
hydrogen.  The problem is that most of the energy produced in the collapse
is lost to the neutrinos that can easily diffuse out of the
newly-born neutron star and stream freely through the infalling matter.
Uncertainties in the complex physics involved in core
collapse -- including strong and weak nuclear processes in
extreme conditions, multidimensional relativistic dynamics,
neutrino transport and probably rotation and  magnetic fields --
have prevented an unambiguous understanding of the explosion mechanism.

The iron (or O/Ne/Mg) core takes about one second to collapse after
instability sets in.  When the neutron star first forms,  a strong
supersonic shock wave propagates back out through the infalling matter.
The halt and rebound of the inner homologous core creates the shock in
about one-hundredth of a second.  As the shock runs out into the
infalling matter, however, some of the energy of the shock is
dissipated by the production and loss of neutrinos.  The shock also
must do the work of breaking down the infalling iron into lighter
elements, protons and neutrons, to form the neutron star.  The shock wave
can thus stall with insufficient energy to reach the outer layers of the
star (Ref. 5).  Matter can continue to rain down on the
stalled shock front.  The matter will still be shocked as material hits
this front, but the shocked matter will continue to fall onto the
neutron star.  If enough matter lands on the neutron star, the neutron
star will be crushed into a black hole.  Current calculations show that
the core bounce and shock formation alone are not sufficient to cause
an explosion (Refs. 143, 170, 171 and 184).
    
One mechanism to create an explosion that is being actively considered
takes advantage of the flux of neutrinos leaving the neutron star
(Ref.  147).  Normal  matter is essentially transparent to
neutrinos since neutrinos  interact only through the weak nuclear force.
Neutron star matter, however, is so dense that it can be opaque or at
least semi-transparent to the neutrinos.  Although most of the neutrinos
will get out, a small fraction will be trapped in the hot matter that
lies just behind the shock front created by the core bounce.  The
accumulation of neutrino heating over a time of about one second may
provide the pressure to re-invigorate the shock, driving the shock
outward and causing the explosion (Ref. 182).
    
The mechanism for depositing a small fraction of the neutrino energy
behind the shock may be related to the convection of the newly-formed
neutron star (Ref. 149).  When the collapse is first
halted and the neutron star rebounds, the composition and thermal
structure are such as to render the proto-neutron star unstable to
convection.  The convection provides a mechanism for carrying the
trapped neutrinos outward by mechanical motion and hence enhancing
the neutrino flux.  Under the right circumstances, this convective
process can be much more efficient in transporting neutrinos than a
slower, essentially diffusive, process.  All modern calculations
that can follow motion in more than one (radial) dimension show that
newly-formed neutron stars are convective (Refs. 158, 143, 170, 171 and
184).    There is a consensus that explosions will not occur without this
convection.   There is still debate about whether this process is sufficient
to cause  an explosion (Refs. 185 and 169).

Most of the current calculations of core bounce and subsequent
events treat the configuration as spherically symmetric.  Even if
the neutron star is convective, the structure of the neutron star
may, on average, be spherically symmetric.  There are a number of
lines of evidence, however, that the explosions that result from
the core-collapse process are intrinsically non-spherical.
Such supernovae are observed to be polarized (Ref. 129).
The pulsars that result from core collapse  have appreciable runaway
velocities (Ref. 118),  implying some sort of asymmetric
``kick" when they are born.  Cas A has evidence for a ``jet" and
perhaps a ``counter jet" of flow (Ref. 70).
Matter may thus be ejected more intensely in some directions than
others.  If this is the case, then the current numerical calculations may be
missing a major ingredient necessary to yield an explosion.  The most
obvious mechanisms for breaking the spherical symmetry are rotation and
magnetic fields.  Proper treatment of rotation and magnetic fields may be
necessary to fully understand when and how collapse leads to explosions.
Recent calculations have suggested that jets of matter associated with
the rotation and magnetic field of the neutron star could play an
important role in causing the explosion (Ref. 176).
This might alter the significance of the convective neutron star and
diminish the effect of the neutrinos.  Alternatively, there may be means of
making the effects of the neutrino flux sufficiently asymmetric
(Refs. 152 and 198).

Theoretical calculations show that heavy elements in reasonable
proportions are produced naturally in massive stars in the process of
forming an iron core (Refs. 202 and 200).
Stars with mass between about 15 and 100 \m\ produce substantial
amounts of heavy elements.  If these stars explode and eject their
heavy elements, this freshly-synthesized material will mix with the
hydrogen in the interstellar gas.  New stars form from this enriched
mixture.  Stars in the upper end of this range could collapse to
form black holes without substantially altering this picture.

Type Ia supernovae are thought to explode in old, low-mass,
long-lived stars.  The observed properties of Type Ia are remarkably,
although not perfectly, uniform.  Since white dwarfs of maximum mass,
known as the Chandrasekhar mass (Ref. 12), would be
essentially identical and hence undergo nearly identical explosions,
the observed homogeneity of Type Ia has pointed to an origin in exploding
white dwarfs.  We now know that all Type Ia supernovae are not exactly
identical, but the basic principle still holds.  The idea is that the
more massive star in an orbiting pair could evolve and form a white dwarf.
The low mass companion could then take a long time to evolve, but the
companion would eventually transfer mass onto the white dwarf
(Ref.  204).   If the total mass accumulated by the white
dwarf approaches the Chandrasekhar  mass of about 1.4 \m, the white dwarf
will then explode (Ref. 133).
    
Careful studies of the observed properties of Type Ia supernovae are
completely consistent with the general picture that the explosion occurs
in a white dwarf.  The information revealed by the evolution of the spectra
is consistent with a configuration in which the denser inner portions of the
exploding star burn all the way to iron, but the outer parts are only
partially burned.  There are two different ways of propagating a
thermonuclear explosion in a white dwarf (Ref. 29).
One is a subsonic burning like a flame, a process called deflagration.
The other is a supersonic burning that is preceded by a shock front,
a process known as detonation.  The most sophisticated current models,
those that best match the data, have the unregulated carbon burning begin
as a turbulent deflagration and then make a transition to a supersonic
detonation (Ref. 173).  Such models naturally create
iron-like matter  in the center and intermediate elements like magnesium,
silicon, sulfur and calcium on the outside.  These models also predict
that the white dwarf is completely destroyed, leaving no compact remnant
like a neutron star or a black hole.  The exact nature of the combustion
is still being explored (Refs. 175, 178, 189, 175 and 197).

Computer models of exploding white dwarfs give results that
match both the observed light curves (Refs. 161 and 194) and this spectral
pattern (Refs. 160 and 162) rather well.  The most successful models adopt
a progenitor that is a  carbon/oxygen white dwarf with a mass very near (but
about 1\% less than) the Chandrasekhar mass (Refs. 160 and 181) even for the
subluminous variety (Refs.
162 and 85) and explodes by
means of a delayed detonation process. This class of model has also had
noted success by predicting spectral features  in the near infrared
(Refs. 203 and 85)  The delayed detonation models also make very
specific predictions for the gamma-ray line emissivity resulting from
radioactive decay, but this aspect has yet to be tested quantitatively
(Ref. 166).  This comparison of theory and observation
thus strongly points to an interpretation of Type Ia supernovae as the
explosion of a carbon/oxygen white dwarf at the Chandrasekhar limit.

Delayed detonation models of exploding Chandrasekhar mass carbon/oxygen
white dwarfs can account for the maximum luminosity/decline rate
correlation (Refs. 162 and 163).
In the models this behavior arises if some stars make the transition
from a subsonic deflagration to a supersonic detonation a little earlier
than in others, and hence make less iron peak elements.  Less
radioactive nickel (see \S I.4) yields a dimmer explosion that is
also cooler with lower opacity so the decline is more rapid.  Why
the conditions of transition should vary is the object of current research.
Recent calculations have given an important insight into the conditions
just prior to the dynamical nuclear runaway.  Carbon first ignites in
these models when the carbon burning rates exceed the neutrino loss
rates (Ref. 133). The carbon burns quasi-statically in a central 
convective core until dynamical runaway ensues.  This convective, 
or ``smoldering" phase, entails convective overturn velocities that 
exceed the predicted velocities of central deflagration waves (Ref. 164).  
This means that the ``pre-conditioning" of the central core will 
be dominated by the convective motions and the subsequent deflagration
phase cannot ``forget" these initial conditions until substantial burning
has ensued in the dynamical phase.  This gives a new way to understand
how similar initial conditions could lead to varying results depending
on just where and how the dynamical runaway began in the convective,
smoldering core.  Rotation and the carbon and oxygen distribution
in the white dwarf (determined by the progenitor star mass) could
also play a role.

Although there is a convergence of opinion on the explosion mechanism,
there is no generally accepted picture of the evolutionary origin of
Type Ia supernovae.  The question of how the white dwarfs grow to the
Chandrasekhar mass is still an unsolved problem.  There is no direct
evidence that Type Ia supernovae arise in binary systems.  Despite this
lack of direct evidence, all the circumstantial evidence points to
evolution in double star systems, and there are few credible ways of
making a white dwarf explode without invoking a binary companion.
The challenge is to figure out what binary evolution leads to a
Type Ia explosion (Refs. 156 and 179).

Type Ia supernovae are thought to eject most of the iron peak
elements in the current interstellar medium, but to have begun
their contribution at a later epoch than the addition of iron from
Type II because of delay in the (uncertain) binary evolution processes
(Refs. 201, 202 and 169).  Type Ia also
eject substantial amounts of O, Mg, Si, S, and Ca.
 
\subsection*{ I.4 Light Curves and Radioactive Decay}
    
Supernovae display a variety of shapes to their light curves
(Ref. 50).  Type Ia supernovae are the brightest.
They decay fairly rapidly in the first two weeks after peak light and
then more slowly for months.  Some Type II supernovae have an extended
plateau and some drop rather quickly from maximum light.  Both types
seem to have a very slow decay at very late times, several months after
the explosion.  Type Ib and Ic supernovae are typically fainter than
Type Ia by about a factor of two, but have similar shapes near peak
light and show evidence for a slow decay at later times.
The common theme to this behavior is radioactive decay.
    
When a supernova first explodes,  the matter is compact, dense and
opaque.  To reach maximum brightness, the ejected matter must expand
until the material becomes more tenuous and semi-transparent.  The
size the ejecta must reach is typically $10^{15}$ cm
(Ref. 150).  The expansion to reach this size results
in adiabatic cooling.  Before reaching this radius, the ejecta may
have cooled sufficiently that there is little heat to radiate.
    
Most Type II supernova explosions are thought to occur in red supergiant
envelopes.  These are very large structures with radii approaching
$10^{14}$ cm.  After the explosion, these large envelopes only
have to expand about an order of magnitude in radius before
they become sufficiently transparent to radiate and hence undergo
relatively little adiabatic cooling.
As they begin to radiate, Type II supernovae
thus still retain a large proportion of the heat that was deposited by
the shock wave that accompanied the supernova.  Near maximum light,
Type II supernovae shine by the shock energy originally deposited in
the star.  In the long, nearly flat ÒplateauÓ phase that characterizes
many Type II, a hydrogen recombination wave propagates inward through
the expanding, cooling envelope (Refs. 150, 154 and 56).   The recombined
hydrogen has low opacity and is nearly transparent.  The still-ionized
hydrogen  is opaque.  The result is that the photosphere sits at the layer
in 
the expanding matter that has a temperature corresponding to the
recombination of hydrogen, about 6000 K, and at a nearly-constant radius,
$\sim10^{15}$ cm, although receding in mass.  The nearly-constant
temperature given by hydrogen recombination and the nearly-constant
radius given by the density gradient and opacity effects, yield the
nearly-constant plateau luminosity until the whole hydrogen envelope has
recombined, at which point the luminosity tends to drop sharply.
    
For a Type I supernova the exploding star is a white dwarf, as suspected
for a Type Ia, or the bare core of a more massive star, as suspected
for Types Ib and Ic.  In either case, the exploding object is very small.
The expected sizes range from $10^{8}$ cm to $10^{10}$ cm.  These bare
cores are thus much smaller than the size to which they must expand
before they can radiate their shock energy.  The result is that the
expansion strongly adiabatically cools the ejected matter and by the
time the matter reaches the point where it could radiate the heat,
the heat from the original shock is dissipated into the kinetic energy
of expansion.  This kind of supernova requires another source of heat to
shine at all.  All the light from Type I supernovae comes
from radioactive decay.
    
The nature of a thermonuclear explosion is to burn very rapidly.  If
the explosion starts with a fuel -- carbon, oxygen, or silicon -- that has
equal numbers of protons and neutrons, then the immediate product of the
burning will also have equal numbers of protons and neutrons.  This is
because the rapid burning takes place on the time scale of the strong
nuclear reactions.   To change the ratio of protons to neutrons requires
the weak force and thus typically a much longer time.
    
Iron is produced in a thermonuclear explosion in a three-step process.
The first step is to produce an element that is close to iron, with the
same atomic weight and a similar large binding energy per nucleon, but
that has equal numbers of protons and neutrons.  This condition singles
out one element: $^{56}$Ni (Refs. 146, 192 and 193).   Nickel-56 is,
however,
unstable to radioactive decay induced by the weak force.  The result is the
formation of the element
$^{56}$Co.  In the process, a positron is emitted to conserve charge, and
a neutrino is given off to balance the number of leptons.  Excess energy
comes off as gamma rays.  The annihilation of the electron and positron will
produce another source of gamma rays.  The gamma rays can be stopped by
collision with the matter being ejected from the supernova and their energy
converted to heat the matter.  The hot matter then radiates.
The resulting optical power falls off as the nickel decays away
with a half-life of 6.1 days and as the matter expands so that it is less
efficient in trapping the gamma rays.  The $^{56}$Co that forms is also
unstable and decays with a half-life of 77 days to become $^{56}$Fe.
This decay again produces a neutrino and gamma-ray energy.
    
The observed light curves of Type I supernovae decay somewhat faster than
the decay of $^{56}$Ni in the early phase and of $^{56}$Co in the
later phases.  The reason is that not all the gamma-rays produced in
the decay are trapped and converted into heat and light.  Some of the
gamma rays escape directly into space.
    
For Types Ib and Ic, the amount of $^{56}$Ni required to power the light
curve is about 0.1 \m\ (Refs. 131 and 148).
This amount of $^{56}$Ni is consistent with many computations of iron-core
collapse.  The $^{56}$Ni is produced when the shock wave impacts the
layer of silicon surrounding the iron core.  Similar reactions occur
in Type II supernovae and similar amounts of $^{56}$Ni are ejected.
The long tail subsequent to the decline from the plateau of Type II
supernovae is consistent with the radioactive decay of this amount of
$^{56}$Co.  Some Type II are observed not to have this tail and this
has been interpreted as signifying the formation of a black hole that
swallows much or all of the $^{56}$Ni that might have been produced in the
shock (Ref. 122).  This hypothesis remains to be proven.

Type Ia supernovae are generally brighter than Type Ib/c and must
produce more $^{56}$Ni, of order 1/2 to 1 \m\ (Ref. 161).
The dimmest Type Ia events require only 0.1 to 0.2 solar masses of $^{56}$Ni
(Ref. 161).  The models of Type Ia supernovae based on
thermonuclear explosions in carbon/oxygen white dwarfs of the
Chandrasekhar mass produce this amount of nickel rather naturally in
the explosion.  The amount of $^{56}$Ni can vary depending on, for
instance, the density at which the explosion makes the transition from a
deflagration to a detonation, so the variety of ejected nickel mass can
also be understood, at least at a rudimentary level (Ref. 161).

\subsection*{ I.5 Supernova 1987A}

Supernova 1987A was a Type II supernovae that occurred in the
Large Magellanic Cloud about 50 kiloparsecs from the Earth
(Refs. 16, 23, 2 and 22). This made it the closest and hence best
studied supernova ever.  Neutrinos were detected from SN 1987A,
thus confirming the basic picture of core collapse (Refs.
61 and 84).  There was insufficient
information to determine the precise mechanism of the explosion.
The neutrinos mean that a neutron star formed, at least temporarily.
No subsequent evidence for a neutron star has been detected.
This may be because the neutron star is obscured in some way, or because it
does not emit the flux of radiation observed from the Crab pulsar.
If SN~1987A produced the same type of compact object as Cas A
(\S I.3), it would not be detectable with current technology.
The failure to detect a compact remnant as a pulsar could be because
the neutron star is spinning rather slowly or has a rather small
magnetic field or is still obscured by matter within the debris.
Another interesting possibility is that sufficient matter fell
back onto the newly-born neutron star to convert it to a black
hole after the supernova shock was launched (Ref. 144).

SN 1987A also gave direct evidence for the production and decay of
about 0.07 solar masses of $^{56}$Ni (Ref. 2).
The gamma-rays from the decay of $^{56}$Co were directly detected.
SN 1987A exploded as a rather compact blue supergiant with a radius
of about $3\times10^{12}$ cm rather than a red giant.  The reasons for
this are still not clear, but current reasoning suggests that the
progenitor might have had a binary companion that was consumed when
the progenitor became a red giant and engulfed the companion
(Ref. 195).  The relatively compact progenitor star meant
that, unlike most Type II, SN 1987A was powered at maximum light by
radioactive decay.  SN 1987A was also the first supernova to have
confirmed production of molecules, particularly carbon-monoxide
(Ref. 116).  Subsequent observations suggest that
production of CO may be rather common in massive-star supernovae
(Ref. 80)
 
SN~1987A is surrounded by a set of three nested rings of gas that
were formed prior to the explosion (Ref. 128).
Their origin is unclear, but may be related to the binary interaction.
The ejecta from the supernova is beginning to collide with the
innermost ring (Refs. 100 and 92).
This interaction  may yield more clues to the origin of this structure.
The explosion of  SN 1987A was clearly asymmetric.  Spectroscopy,
spectropolarimetry, and  direct imaging suggest a common axis roughly
(but not precisely) aligned  with the axis normal to the outer ring
systems (Ref. 172).

\subsection*{ I.6 Supernovae, Gamma-ray Bursts, and Cosmology}
    
Because they are so bright, supernovae have been used to determine
distances.  This information can be combined with the expansion
velocity of the host galaxy to determine the Hubble Constant and hence the
age of the Universe.  Type Ia supernovae have proven to be especially
powerful tools in this way.  Subtle differences in the brightness
for more distant Type Ia supernovae have given evidence that the
Universe is accelerating its expansion, contrary to expectation
(Refs. 113 and 106).  Taken at
face value and in combination with other data on the cosmic microwave
background radiation, this evidence suggests that the Universe
is filled with a ``dark energy" that adds a repulsive component
to the dynamics of the Universe (Ref. 136).  This
dark energy could be related to Einstein's Cosmological Constant if
constant in space or time, or to some form of vacuum energy that
could vary in space and time.  This anti-gravitating dark energy is
to be contrasted with dark matter, which, although of an as yet
unknown constituency, gravitates.

After the Universe expanded and hydrogen recombined, the era of
the creation of the cosmic background radiation, the matter in
the Universe was dark and nearly formless.  Over the eons of expansion,
the matter coalesced into structures that we identify today as stars,
galaxies, galaxy clusters and superclusters.  At some  point
in the formation of this structure, the first stars were born.
This epoch in the evolution of the Universe is known as the end of
the Dark Ages that separated recombination from the first stars
(Ref. 205).  The end of the Dark Ages will probably be
marked by the explosion of the first supernovae that are likely to be
among the brightest objects in the Universe at this epoch
(Ref. 186).  These supernovae should be some form
of Type II, but they may be especially massive and they may be
relatively compact since they will be formed from pristine matter not
yet polluted with heavy elements.  The resulting low-opacity envelopes
may selectively produce blue rather than red supergiants
(Ref. 165).  The Next Generation Space Telescope may be
able to detect these first supernovae and hence to map the evolution
of the production of supernovae, of the heavy elements they produced,
radiation that re-ionized the matter between galaxies, and the onset
of the first Type Ia supernovae.
    
There has been a revolution in the study of cosmic gamma-ray bursts
with the discovery of optical counterparts and proof that the large
majority of these events are at large redshift (Ref. 127).
Growing evidence suggests that gamma-ray bursts are collimated explosions
with energies of about $10^{51}$ ergs associated with the death
of massive stars (Refs. 191 and 77).
Essentially all gamma-ray bursts identified with host galaxies are within
the optical confines of those galaxies, which show evidence for recent
star formation (Ref. 86).  More specifically, two
gamma-ray bursts have been associated with supernova-like brightening
several weeks after the explosion (Refs. 112 and 62).   SN 1998bw had
characteristics of a Type Ic  supernova, but was excessively bright,
asymmetric, had large expansion  velocities, and produced one of the
brightest
radio sources ever  associated with a supernova (Ref. 78).
There are  indications that this supernova was associated with a gamma-ray
burst  observed on April 25, 1998.  If this association is correct, this
especially-bright supernova was associated with an especially dim
gamma-ray burst.  The association of gamma-ray bursts with massive
stars suggests that gamma-ray bursts represent some variation of the
physics of core collapse.  A popular hypothesis is that gamma-ray
bursts represent the birth of black holes from core collapse
(Ref. 37), but there is  no direct evidence of this and
the energetics do not preclude a role for neutron star formation.
The asymmetries observed  in core collapse supernovae may have some
generic connection  to the collimation deduced for the gamma-ray
bursts.   Future space missions, for instance the Swift explorer,
have the promise of detecting gamma-ray bursts at high  redshifts
giving another way to probe the death of massive stars at the end of
the Dark Ages when the first stars begin to be born and die.

\section*{II. JOURNALS}

{\it
Annual Reviews of Astronomy and Astrophysics\\
Astronomical Journal\\
Astronomy and Astrophysics\\
Astrophysical Journal\\
Astrophysical Journal Letters\\
Astrophysical Journal Supplemant\\
Astrophysics and Space Science\\
Memoirs della Societa Astronomica Italia\\
Monthly Notices of the Royal Astronomical Society\\
Nature\\
Nuclear Physics A\\
Physica Scripta\\
Physical Review D\\
Physical Review Letters\\
Publications of the Astronomical Society of the Pacific\\
Revista Mexicana de Astronomia y Astrofysica\\
Science\\
Soviet Astronomy
}

\section*{III. BOOKS AND REVIEW ARTICLES}

\begin{enumerate}
\item
{\bf Supernovae and Nucleosynthesis}, W.~D. Arnett (Princeton
University Press, Princeton, 1996).  This textbook concentrates mostly on
the stellar evolution and nucleosynthesis that precede explosions and the
nucleosynthesis the transpires in the explosion. (I, A)

\item
 ``Supernova 1987A," W.~D. Arnett, J.~N.  Bahcall,
 R.~P. Kirshner \&  S.~E. Woosley,\, Ann. Rev. Astron. Astrophys.,
 \textbf{27}, 629-700 (1989). This is the first comprehensive review of the
flood of work done on SN~1987A. (I, A)

\item
``Neutrino Processes in Stellar Interiors," Z. Barkat, Ann. Rev. Astron.
Astrophys.,
\textbf{13}, 45-68 (1975).  Review of neutrino processes in  supernovae and
their
progenitors. (I, A)

\item
{\bf Supernovae as Distance Indicators}, edited by N. Bartel
(Springer-Verlag, New York, 1985).  Discussion of the observations
and theory of supernovae with the theme of determining the distance
scale.  Early work on using limits on the ejection of radioactive
$^{56}$Ni to constrain peak luminosity of Type Ia. (E, I, A)

\item
``Supernova Mechanisms," H. A. Bethe, Rev. Mod. Phys., {\bf 62}, 801-867
(1990).  
Thorough discussion of the physics of core collapse, neutrino trapping, and
neutron
star convection. (E, I, A)

\item
{\bf Supernovae}, edited by S. A. Bludman, R. Mochkovitch \& J. Zinn-Justin
(Elsevier, Amsterdam, 1994).  Complilation of lectures from summer school.
Noted for
review of  models of thermonuclear and core collapse explosion and
associated
nucleosynthesis. (I, A)

\item
``The Physics of Supernovae," edited by S. Bludman D. H. Feng, T. Gaisser,
\& S.
Pittel,  Phys. Rep., {\bf 256}, 1-235 (1995).  Symposium honoring the
presentation of
the Wetherill Medal of the Franklin Institute to Stirling Colgate.  Topics
include
core collapse, binary star explosions and models for Type Ia supernovae..
(I. A)

\item
``Type Ia Supernovae as Standard Candles," D. Branch,~\& G.~A. Tammann, Ann.
Rev. Astron. Astrophys,
\textbf{30}, 359-389 (1992).  This review summarized rates of supernovae and
implications for the Hubble Constant. (I, A)

\item
``Type IA Supernovae and the Hubble Constant,"  D. Branch, Ann. Rev. Astron.
Astrophys., \textbf{36}, 17-56 (1998).  This review summarizes spectra of
supernovae and implications for the Hubble Constant. (I, A)

\item
``Theory of Supernovae," edited by G. E. Brown, Phys. Rep., \textbf{163}
1-204
(1988).
Proceedings of Symposium honoring Hans Bethe's 80th birthday.  Reviews of
evolution
and explosion of massive stars noted for presentation of nucleosynthesis by
Nomoto
and Hashimoto and of core-collapse dynamics by Wilson and Mayle. (E, I, A)

\item
{\bf Future Directions of Supernova Research: Progenitors to
Remnant}, edited by S. Cassisi \& P. Mazzali (Societ\'a
Astronomical Italiana, Rome, 2000).  Discussion of observations
and theory of supernovae.  Up-to-date discussion of degenerate
core progenitor evolution. (E, I, A)

\item
\textbf{An Introduction to the Study of Stellar Structure}, S. Chandrasekhar
(University of Chicago Press, Chicago, 1939).  Classic discussion
of the nature of the electron degenerate equation of state and the structure
of white dwarfs and their eponymous limiting mass. (E, I, A)

\item
``The Interaction of Supernovae with the Interstellar Medium," R.~A.
Chevalier, Ann.
Rev. Astron. Astrophys.,
\textbf{15}, 175-196 (1977).  Review of supernova remnant formation. (I, A)

\item
\textbf{The Historical Supernovae}, D.~H. Clark, \&  F.~R. Stephenson,
(Pergamon Press, Oxford, 1977). Classic presentation of observations of the
historical record from Chinese records onward. (E, I)

\item
{\bf Supernovae and Supernova Remnants}, edited by C. B. Cosmovici
(D. Reidel, Dordrecht, 1974).  One of the early comprehensive
collections on supernovae, see especially the classic analysis of
supernovae statistics by Tammann. (E, I, A)

\item
{\bf Workshop on SN 1987A}, edited by I. J. Danziger (ESO, Garching, 1987).
First workshop with early results on SN 1987A. (E, I, A)
 
\item
``Optical Spectra of Supernovae," A.~V. Fillipenko, Ann. Rev. Astron.
Astrophys., \textbf{35}, 309-355 (1997).
Recent review of supernova spectra. (I, A)

\item
``Classification of Supernovae," R. P. Harkness \& J. C. Wheeler,
in  {\bf Supernovae}, edited by
A. G. Petschek (Springer-Verlag, New York, 1990), pp. 1 - 29. 
Early review placing Type Ib and Ic in the spectral classification
scheme. (I, A)
 
\item
``Type Ia Supernova Explosion Models,"  W. Hillebrandt,~\& J.~C. Niemeyer,
Ann. Rev. Astron. Astrophys.,
 \textbf{38}, 191-230 (2000). Review of progress toward understanding the
thermonuclear
physics of Type Ia explosions. (I, A)

\item
{\bf Cosmic Explosions}, edited by S. S. Holt and W. W. Zhang
(American Institute of Physics, New York, 2000).  Discussion of
the supernova/gamma-ray burst connection and related issues. (E, I, A)

\item 
``The Terminal Phases of Stellar Evolution and the Supernova Phenomenon,"
V. S. Imshennik \& D. K. Nadyozhin,  Sov. Sci. Rev., {\bf 2},
75-161 (1983). Early review of supernovae by two pioneers of Russian
supernova
research. (I, A)

\item
``Supernova 1987A,"  V. S. Imshennik, \& D. K. Nadyozhin,
Sov. Sci. Rev., E {\bf 8}, 1-143 (1989). Review
of SN~1987A.  (I, A)

\item
{\bf Supernova 1987A in the Large Magellanic Cloud}, edited
by M. Kafatos and A. Michalitsianos (Cambridge University Press,
Cambridge, 1988).  Compilation of work in the first heady year
after the advent of SN~1987A. Features discussions of the detection
of neutrinos, polarization, atmosphere models. (E, I, A)

\item
``Thermonuclear Burning and The Explosion of Degenerate Matter in
Supernova," A. M. Khokhlov, Sov. Sci. Rev., E \textbf{8} (2), 1-75 (1989).
Review of the physics of thermonuclear explosions.  (I, A)

\item
``Cosmological Implications from Observations of Type Ia
Supernovae,"  B. Leibundgut,
Ann. Rev. Astron. Astrophys., \textbf{39}, 67-98 (2001). Recent review of
the
nature
of Type Ia supernovae and the implications for an accelerating Universe. (I,
A)  

\item
{\bf Supernovae and Gamma-Ray Bursts: The Greatest Explosions
Since the Big Bang}, edited by M. Livio, N. Panagia, and K. Sahu
(Cambridge University Press, Cambridge, 2001).  Discussions of
the accelerating Universe as revealed by Type Ia supernovae
and evidence of a supernova/gamma-ray burst connection. (E, I, A)

\item
``Supernova 1987A Revisited," R. McCray, Ann. Rev. Astron. Astrophys.,
\textbf{31}, 175-216 (1993).  Follow-up review of the next several year's
work
on SN~1987A,
especially as it entered the nebular phase. (I, A)

\item
{\bf Supernovae and Supernova Remnants}, edited by R.  A. McCray, \&
Z. Wang, (Cambridge University Press, Cambridge, 1996). Reviews of
supernovae
including IR spectra of SN 1987A by Worden and of SN 1993J by
Wheeler and Filippenko.
(E, I, A)

\item
``Thermonuclear Explosions in Stars,"  T. J. Mazurek \& J. C. Wheeler,
Fund. Cosmic Phys., {\bf 5}, 193-286 (1980).  Review of thermonuclear
physics
relevant to Type Ia explosions, detonation and deflagration physics. (I, A)

\item
``Supernovae and Supernovae Remnants," R. Minkowski, Ann. Rev. Astron.
Astrophys.,
\textbf{2}, 247-266 (1964).  Classic early review of the nature of
supernovae,
Types I and II. (E, I)

\item
``The Spectra of Supernovae," J.~B. Oke,~\&  L. Searle, Ann. Rev. Astron.
Astrophys.,
\textbf{12}, 315 (1974).  Classic early review of spectral classifications
and physics. (E, I)

\item
{\bf Supernovae}, edited by A. G. Petschek (Springer-Verlag, New York,
1990).
Complilation of invited review papers.  Noted for the article
by Harkness and Wheeler, (Ref. 18 )
that presented the ``tree diagram" of
the increasingly complicated spectra classification scheme that
now included Types Ib and Ic. (E, I, A)
  
\item
{\bf Supernovae: A Survey of Current Research}, edited by
M. J. Rees and R. J. Stoneham (Reidel, Dordrecht, 1982).
Comprensive survey of contemporary work on observations and
theory of supernovae, remnants and applications to the distance scale.
(E, I, A)

\item
{\bf Thermonuclear Supernovae}, edited by P. Ruiz-Lapuente, R. Canal,
and J. Isern (Kluwer, Dordrecht, 1997).  Proceedings of a meeting
that exhaustively presented the current picture of Type Ia supernovae,
their observations and physics and implications for the distance scale.
First discussion of specific mechanisms of deflagration to detonation
transition by Khokhlov et al. and the beginning of the end of
sub-Chandrasekhar models for Type Ia. (E, I, A)
 
\item
{\bf Supernovae}, edited by D. N. Schramm (Reidel, Dordrecht, 1977).
Useful cross section of work on observations and theory of supernovae.
(E, I, A)

\item
{\bf Explosive Nucleosynthesis}, edited by D. N. Schramm and W. D. Arnett
(University of Texas Press, Austin, 1973).
Discussions of explosion mechanisms with emphasis on the associated
nucleosynthesis. (E, I, A)

\item
``Nuclear Astrophysics," edited by D. N. Schramm \& S. E. Woosley,
Physics Reports, {\bf 227}, 1-319 (1993).  Festschrift in honor of
Willy Fowler's 80th Birthday, including a broad summary of topics relevant
to supernovae and nucleosynthesis. (E, I, A)

\item
{\bf Supernovae}, I. S. Shklovskii (Wiley-Interscience, New York, 1968).
First book on supernovae including classic discussion of Crab Nebula.
(E I)

\item
``Presupernova Models and Supernovae," D. Sugimoto, and K. Nomoto, Space
Sci. Rev., \textbf{25}, 155-227 (1980). Review of work on degenerate
supernova progenitors. (I, A)

\item
``What Stars Become Supernovae?" B. M. Tinsley, Pub. Ast. Soc. Pac.,
{\bf 87}, 837-848 (1975). Classic review of the origin of supernovae. (E. I)

\item
``Supernovae Part I: The Events," V. Trimble, Rev. Mod. Phys., {\bf 54},
1183-1224 (1982).  Review of the history of supernova research, progenitor
evolution,
observations, and models.  (E, I)

\item
``Supernovae Part II: The Aftermath," V. Trimble, Rev. Mod.
Phys., {\bf 55}, 511-562 (1982).  Review of the evolution of
supernovae into supernova remnants,
compact objects and nucleosynthesis.  (E, I)

\item ``Galactic and Extragalactic
Supernova Rates," S. van den Bergh,~\& G.~A. Tammann, Ann. Rev. Astr.
Astrophys.,
\textbf{29}, 363-407 (1991).  Classic discussion of supernova
statistics and rates with occasionally departing points of view between the
authors. (I, A)

\item
``Radio Emission from SNe and Young SNRs," K. W. Weiler, N. Panagia,
M. J. Montes, S. D. van Dyk, R. A. Sramek \& C. K. Lacey,
in {\bf Young Supernova Remnants}, edited by S. S. Holt \& U. Hwang
(American Institute of Physics, Melville, NY, 2001) pp. 237-246. 
Recent review of radio properties of supernovae.  (E. I)

\item
``Supernovae and Supernova Remnants," K.~W. Weiler,~\& R.~A. Sramek,
Ann. Rev. Astr. and Astrophys., \textbf{26}, 295 (1988).
Early review of radio properties of supernovae. (I, A)

\item
``Supernovae in Binary Systems," J. C. Wheeler, in {\bf Frontiers of
Stellar Evolution}, edited by D. L. Lambert (Astr. Soc. Pac., San Francisco,
CA, 1991), pp. 483-538.  Overview of binary evolution pertaining to
supernovae.  One of first discussions of potential importance of detecting
circumstellar environment of Type Ia. (I, A)

\item
{\bf Supernovae}, edited by J. C. Wheeler, T. Piran \& S. Weinberg
(World Scientific, Singapore, 1990).
Invited reviews from a Winter School.  Noted for the excellent review
of line formation and transfer in supernovae by Jeffery and Branch
and perhaps the first publication of work on non-monotonic massive
star progenitor core evolution by Barkat and Marom. (E, I, A)

\item
{\bf Proceedings of the Texas Workshop on Type Ia Supernovae},
edited by J. C. Wheeler (University of Texas, Austin, 1980).
Workshop proceedings known for pioneering work of Axelrod on
the radioactive excitation of nebular spectrum of Type Ia supernovae
and for early discussions of ``off-center" ignition models
by Woosley, Weaver, and Taam. (E, I, A)

\item
``The Origin of Supernovae," J. C. Wheeler, Rep. Prog. Phys., {\bf 44},
89-138 (1981).  Review of pre-supernova evolution and explosion mechanisms.
(E, I)

\item
``Supernovae," J. C. Wheeler \& S. Benetti, in {\bf Allen's Astrophysical
Quantities}, edited by A. N. Cox (Springer Verlag, New York, 2000),
pp. 451-469. Compact review of supernova properties. (E. I)

\item 
``Review of Contributions to the Workshop on SN 1993J," J. C. Wheeler \&
A. V. Filippenko, in {\bf Supernovae and Supernova Remnants}, edited by R.
A. McCray, Z. Wang, \& Z. Li (Cambridge University Press, Cambridge, 1995),
pp. 241-276. Compilation and analysis of observations and theory of
SN 1993J. (I, A)
 
\item
``Type I Supernovae," J. C. Wheeler \& R. P. Harkness, Rep. Prog. Phys.,
\textbf{53}, 1467-1557 (1990). Review of the observational and theoretical
properties of Type I supernovae. (E, I)

\item
``Abundance Ratios as a Function of Metallicity," J. C. Wheeler,
C. Sneden \& J. W. Truran, Jr. Ann. Rev. Astr. Astrophys., {\bf 27}, 279-349
(1989).  Review of stellar abundances and nucleosynthetic history. (E, I)

\item
``Supernova Remnants," L. Woltjer, Ann. Rev. Astron. Astrophys.,
\textbf{10},
129-159
(1972).  Classic early review of the properties of supernovae. (E, I)

\item
{\bf Supernovae}, edited by S. E. Woosley (Springer-Verlag, New York, 1991).
Comprehensive collection of work on SN 1987A.  Known for the
papers on dust formation by Lucy, Danziger, and colleagues.
(E, I, A) 

\item
``The Physics of Supernova Explosions," S.~E. Woosley,~\& T.~A. Weaver,
Ann. Rev. Astron. Astrophys., \textbf{24}, 205-253 (1986). Review of models
of
thermonuclear and core-collapse explosion and associated nucleosynthesis.
(I,
A)

\section*{IV. ELABORATIONS AND APPLICATIONS}

\subsection*{ IV.1 Observations}

\item
``Photographic Light-Curves of the Two Supernovae in IC 4182 and NGC
1003," W. Baade,
\& F. Zwicky, Astrophys. J., \textbf{88}, 411-422 (1938). (E, I)

\item
``Type Ia Supernova 1989B in NGC 3627," R. Barbon, S. Benetti, E.
Cappellaro,
 L. Rosino,
\& M. Turatto, Ast. \& Astrophys., {\bf 237}, 79-90 (1990). (I, A)

\item
``The Asiago Supernova Catalogue - 10 years After," R. Barbon, V. Buond',
E. Cappellaro \& M. Turatto, Ast. \& Astrophys. Supp., {\bf 139}, 531-536
(1999).  See also 
http://simbad.u$\sim$strasbg.fr/cgi-bin/cdsbib?1999A\%26AS..139..531B
and \\ http://www.RochesterAstronomy.org/\\supernova.htm.
(I, A) 

\item
``Distribution of Supernovae Relative to Spiral Arms and H II Regions," O.
S. Bartunov, D. Y. Tsvetkov, and I. V. Filimonova, Pub. Astron. Soc. Pac.,
{\bf 106}, 1276-1284 (1994).  (I, A)

\item
``Observation of a Neutrino Burst in Coincidence with Supernova 1987A
in the Large Magellanic Cloud," R. M. Bionta, G. Blewitt, C. B. Bratton,
D. Caspere, and A. Ciocio, Phys. Rev. Lett., {\bf 58}, 1494-1496
(1987). (I, A)

\item
``The Unusual Afterglow of the Gamma-Ray Burst of 26 March 1998 as Evidence
for a Supernova Connection," J. S. Bloom, S. R. Kulkarni, S. G. Djorgovski,
A. C. Eichelberger, P. Cote, J. P. Blakeslee, S. C. Odewahn, F. A. Harrison,
D. A. Frail, A. V. Filippenko, D. C. Leonard, A. G. Riess, H. Spinrad,
D. Stern, A. Bunker, A. Dey, A.,  B. Grossan, S. Perlmutter, R. A. Knop,
I. M. Hook, M. Feroci, Nature, {\bf 401}, 453-456 (1999). (I, A)

\item
``The Spectrum of the Type I Supernova of 1954 in NGC 4214," D. Branch,
Astron. \& Astrophys., {\bf 16}, 247-252 (1972). (E, I)

\item
``The Type II Supernovae 1979c in M100 and the Distance to the Virgo
Cluster,"
D. Branch, S. W. Falk, M. L. McCall, P. Rybski, A. K. Uomoto, \& B. J.
Wills,
Astrophys. J., {\bf 244}, 780-804 (1981). (I, A)

\item
``The Type I Supernovae 1981b in NGC 4536 - The First 100 Days," D. Branch,
C. H.
Lacy, M. L. McCall, P. G. Sutherland, A. Uomoto, J. C.  Wheeler, \& B. J.
Wills,
Astrophys. J., {\bf 270}, 123-129 (1983). (I, A)

\item
`` The Rate of Supernovae from the Combined Sample of Five Searches," E.
Cappellaro,
M. Turatto, D. Yu. Tsvetkov, O. S. Bartunov, C. Pollas, R. Evans, \& M.
Hamuy, Astron.  \& Astrophys., \textbf{322}, 431-441 (1997). (I, A)

\item
``The Type II Supernova 1969 I in NGC 1058," F. Ciatti, L. Rosino, \& F.
Bertola, 
Mem. Soc. Astron. Italia, {\bf 42}, 163 (1971). (I, A)

\item
``Type I Supernovae in the Infrared and their use as Distance Indicators,"
J. H. 
Elias, K. Matthews, G. Neugebauer, and S. R. Persson, Astrophys. J., {\bf
296},
379-389 (1985). (I, A)

\item
``Optical and Infrared Spectroscopy of the Type IIn SN 1998S: Days 3-127,"
A. Fassia, W. P. S. Meikle, N. Chugai, T. R. Geballe, P.
Lundqvist, N. A. Walton, D. Pollacco, S. Veilleux, G. S.
Wright, M. Pettini, T. Kerr, E. Puchnarewics, P. Puxley, M.
Irwin, C. Packham, S. J. Smartt \& D. Harmer,  Mon. Not. Roy.
Astr. Soc.,
\textbf{325}, 907-930 (2001). (I, A)

\item
``An Optical Survey of Outlying Ejecta in Cassiopeia A: Evidence for a
Turbulent, Assymetric Explosion," R. A. Fesen, Astrophys. J. Supp.,
\textbf{133},
161 - 186 (2001). (I, A)

\item 
``Optical Spectroscopy and Imaging of the Northeast Jet in the Cassiopeia A
Supernova Remnant," R. A. Fesen,
\& K. S. Gunderson, Astrophys. J., \textbf{470}, 967-980 (1996). (I, A)

\item 
``Supernova 1987K - Type II in Youth, Type Ib in Old Age," A. V. Filippenko,
Astrophys. J., {\bf 96}, 1941-1948 (1988). (I, A)

\item 
``The `Seyfert 1' Optical Spectra of the Type II Supernovae 1987F and
1988I,"
 A. V. Filippenko, Astrophys. J., {\bf 97}, 726-734 (1998). (I, A)

\item
``The Type Ic (Helium Poor) Supernova 1987M - Transition to the
Supernebular Phase,"  A. V. Filippenko, A. C. Porter,
\& W. L. W. Sargent, Astrophys. J., {\bf 100}, 1575-1587 (1990). (E, I)

\item 
``The Peculiar Type Ia SN 1991T - Detonation of a White Dwarf?" A. V.
Filippenko, M. W. Richmond, T. Matheson, J. C. Shields, E. M. Burbidge, R.
D.
Cohen, M. Dickinson, M. A. Malkan, B. Nelson, J. Pietz, D. Schlegel, P.
Schmeer,
H. Spinrad, C. C. Steidel, H. D. Tran, W. Wren, Astrophys.
J. Lett., {\bf 384}, L15-L18 (1992). (I, A)

\item
``The Subluminous, Spectroscopically Peculiar Type Ia
Supernova 1991bg in the Elliptical Galaxy," A. V. Filippenko, M. W.
Richmond,
D. Branch, M. Gaskell, W. Herbst, C. H. Ford, R. R. Treffers, T. Matheson,
L.
C. Ho, A. Dey, W. L. W. Sargent, T. A. Small, \& W. J. M. van Breugel,
Astrophys. J., {\bf 104}, 1543-1556 (1992). (I, A)

\item
``Beaming in Gamma-Ray Bursts: Evidence for a Standard Energy Reservoir,"
D. A. Frail, S. R. Kulkarni, R. Sari, S. G. Djorgovski, J. S. Bloom,
T. J. Galama, D. E. Reichart, E. Berger, F. A. Harrison, P. A. Price,
S. A. Yost, A. Diercks, R. W. Goodrich, R. W. \& F. Chaffee, Astrophys. J.,
{\bf 562}, L55-L58 (2001). (I, A)

\item
``An Unusual Supernova in the Error Box of the Gamma - Ray Burst of
25 April 1998," T. J. Galama, ÊP.ÊM. Vreeswijk, J. vanÊParadijs,ÊC.
Kouveliotou, T. Augusteijn, H. Bohnhardt,ÊJ.ÊP. Brewer,ÊV. Doublier,ÊJ.-F.
Gonzalez,ÊB. Leibundgut,ÊC. Lidman,ÊO.ÊR. Hainaut, F. Patat, J. Heise,
in't J. Zand, K. Hurley, P. J. Groot, R. G. Strom, P. A. Mazzali, K.
Iwamoto, K. Nomoto, H. Umeda, T. Nakamura, T. R. Young, T. Suzuki,
T. Shigeyama, T. Koshut, M. Kippen, C. Robinson, P. deÊWildt, R. A. M. J.
Wijers, N. Tanvir, J. Greiner, E. Pian, E. Palazzi, F. Frontera, N. Masetti,
L. Nicastro, M. Feroci, E. Costa, L.  Piro, B. A. Peterson, C. Tinney,
B. Boyle, R. Cannon, R. Stathakis, E. Sadler, M. C. Begam, \& P. Ianna,
Nature, {\bf 395}, 670-672 (1998). (I, A)

\item
``Type Ib Supernovae 1983n and 1985f - Oxygen-rich Late Time Spectra," C. M.
Gaskell,
E. Cappellaro, H. L. Dinerstein, D. R. Garnett, R. P. Harkness, \& J.
C. Wheeler,
Astrophys. J., {\bf 306}, L77-L80 (1986). (I, A)

\item
``Detection of CO and Dust Emission in Near-Infrared Spectra of SN 1998S,"
C. L. Gerardy, R. A. Fesen, P. H\"oflich \& J. C. Wheeler, Astron. J.,
\textbf{119}, 2968-2981, (2000). (I, A)

\item 
``Limitations Imposed on Statistical Studies of
Galactic Supernova Remnants by
Observational Selection Effects," D. A. Green,
Pub. Astr. Soc. Pac., {\bf 103}, 209-220 (1991).  See also:\newline
http://www.phy.cm/ac.uk/www/research/ra/SNRs\\/snrs.intro.html;DD\newline
and\newline
http://hea$\sim$www.harvard.edu/$\sim$slane/snr/cats/\\einstein$\_$list/snr$
\_$doc.html.
(E, I)

\item 
``The Early Spectral Phase of Type Ib Supernovae - Evidence for Helium," R.
P. Harkness, J. C.  Wheeler, B. Margon, R. A. Downes,
R. P. Kirshner, A. Uomoto, E. S. Barker, A. L.  Cochran,
H. L. Dinerstein, D. R. Garnett \& R. M. Leverault,
Astrophys. J., {\bf 317}, 355-367 (1987). (I, A)

\item
``An Early-time Infrared and Optical Study of the Type Ia Supernova
1998bu in M96," M. Hernandez, W. P. S. Meikle, A. Aparicio,
C. R. Benn, M. R. Burleigh, A. C. Chrysostomou, A. J. L.  Fernandez, T.
R. Geballe, P. L. Hemmersley, J. Iglesias-Paramo, D. J. James, P. A. James,
S. N. Kemp, T. A. Lister, D. Martinez-Delgado, A. Oscoz, D. L. Pllacco, M.
Rozas, S. J. Smartt, P. Sorensen, R. A. Swaters, J. H. Telting,
W. D. Vacca, N. A. Walton \& M. R. Zapatero-Osorio, Mon. Not. Roy. Astr.
Soc.,
\textbf{319}, 223-234 (2000). (I, A)

\item
``Observation in the Kamiokande-II Detector of the Neutrino Burst
from Supernova SN1987A," K. S. Hirata, T. Kjita, M. Koshiba, M. Nakahata, Y.
Oyama, N. Sato, A. Suzuki, M. Takita, Y. Totsuka, T. Kifune, T. Suda, K.
Takahashi, T. Tanimori, K. Miyano, M. Yamada, E. W. Beier, L. R. Feldscher,
W.  Frati, S. B. Kim, A. K. Mann, F. M. Newcomer, R. van Berg,
W. Zhang \& B. G.  Cortez, Phys. Rev. D., {\bf 38}, 448-458 (1988). (I, A)

\item
``Infrared Spectra of the Subluminous Type Ia Supernova 1999by,"
P. H\"oflich, C. Gerardy, R. Fesen \& S. Sakai, Astrophys. J., \textbf{568},
791-806 (2002). (I, A)

\item
``The Faint-Galaxy Hosts of Gamma-Ray Bursts," D. W. Hogg \& A. S. Fruchter,
Astrophys. J., {\bf 520}, 54-58 (1999). (I, A)

\item
``The Progenitors of Subluminous Type Ia Supernovae," D. A. Howell,
Astrophys. J., Lett., \textbf{554}, L193-L196 (2001). (I, A)

\item
``Evidence for Asphericity in a Subluminous Type Ia Supernova:
Spectropolarimetry of SN 1999by," D. A. Howell, P. H\"oflich,
L. Wang \& J. C. Wheeler, Astrophys. J., \textbf{556}, 302-321 (2001). (A)

\item
``Supernova 1972e in NGC 5253," R. P. Kirshner, and J. B. Oke, Astrophys. J.
\textbf{200},
574-581 (1975).(I, A)

\item 
``The Spectra of SN," R. P. Kirshner, J. B. Oke, M. V. Penston, \& L.
Searle, 
Astrophys. J., {\bf 185}, 303-322 (1973). (I, A)

\item
``SN 1992A: Ultraviolet and Optical Studies Based on HST, IUE, and CTO
Observations," R. P. Kirshner, D. J. Jeffery, B. Leibundgut,
P. M. Challis, G. Sonneborn, M. M. Phillips, N. B. Suntzeff,
R. C. Smith, P. F. Winkler, C. Winge, M. Hamuy,
D. A. Hunter, K. C. Roth,ÊJ. C. Blades, D. Branch,
R. A. Chevalier, C. Fransson, N. Panagia, R. V. Wagoner, J. C.
Wheeler, \& R. P. Harkness, Astrophys. J., {\bf 415}, 589-615 (1993).
(I, A)

\item
``On the Emergence and Discovery of Hot Spots in SNR 1987A," S. S. Lawrence,
B. E. Sugerman, P. Bouchet, A. P. S. Crotts, R. Uglesich \& S. Heathcote,
Astrophys. J. Lett., {\bf 537}, L123-L126 (2000). (I. A)

\item
``SN 1991bg - A Type Ia Supernova with a Difference," B. Leibundgut,
R. P. Kirshner, M. M. Phillips, L. A. Wells, N. B. Suntzeff,
M. Hamuy, R. A. Schommer, A. R. Walker, L. Gonzalez, P. Ugaarte,
R. E. Williams, G. Williger, M. Gomez, R. Marzke,
B. P. Schmidt, B. Whitney, N. Coldwell, J. Peters,
F. H. Chaffee, C. B. Foltz, D. Rehner, L. Siciliano, T. G. Barnes,
K.-P. Cheng, P. M. N. Hintzen, Y.-C. Kim, J. Maza, J. Wm. Parker,
A. C. Porter, P. C. Schmidtke \& G. Sonneborn, Astron.  J., {\bf 105},
301-313 (1993). (I, A)

\item
``Is It Round? Spectropolarimetry of the Type II-p Supernova 1999EM,"
D. C. Leonard, A. V. Filippenko, D. R. Avila \& M. S. Brotherton,
Astrophys J., {\bf 553}, 861-885 (2001). (A)

\item
``Optical Observations of Supernova 1993J from La-Palma - Part One - Days 2
To 125," J. R. Lewis, N. A. Walton, W. P. S. Meikle, R. J.
Cumming, R. M. Catchpole, M. Arevalo, R. W. Argyle, C. R. Benn,
P. S. Bunclark, H. O. Castaneda, M. Centurion, R. E. S. Clegg,
A. Delgado, V. S. Dhillon, P. Goudfrooij, E. H. Harlaftis,
B. J. M. Hassall, L. Helmer, P. W. Hill, D. H. P. Jones,
D. L. King, C. Lazaro, J. R. Lucey, E. L. Martin, L. Miller,
L. V. Morrison, A. J. Penny, E. Perez, M. Read, P. J. Rudd,
R. G. M. Rutten, R. M. Sharples, S. W. Unger \& J. Vilchez, Mon. Not.
Roy. Astr. Soc.,
\textbf{266}, L27-L39 (1994). (I, A)

\item
``The Parkes Southern Pulsar Survey - II. Final Results and Population
Analysis," A. G. Lyne, R. N. Manchester, D. R. Lorimer, M. Bailes,
N. D'Amico, T. M. Tauris, S. Johnston, J. F. Bell \& L. Nicastro,
Mon. Not. Royal. Astr. Soc., {\bf 295}, 743-755 (1998). (I, A)

\item
``Statistics of Extragalactic Supernovae," J. Maza and  S. van den Bergh,
Astrophys. J., {\bf 204}, 519-529 (1976).(E, I)

\item
``Constraining the Ages of Supernova Progenitors. I. Supernovae and Spiral
Arms," R. J. McMillan \& R. Ciardullo, Astrophys. J., {\bf 473}, 707-712
(1996). (I, A)

\item
``Spectroscopy of Supernova 1987A at 1-5 Microns. I - The First Year,"
W. P. S. Meikle, J. Spyromilio, G.-F. Varani, \& D. A. Allen, Mon. Not.
Roy. Astr. Soc., \textbf{238}, 193-223 (1989). (I, A)

\item
``New Hubble Space Telescope Observations of High-Velocity Ly$\alpha$
and H$\alpha$ in SNR 1987A," E. Michael, R. McCray, C. S. J. Pun, K.
Borkowski, 
P. Garnavich, P. Challis, R. P. Kirshner, R. Chevalier, A. V. Filippenko,
C. Fransson, N. Panagia, M. Phillips, B. Schmidt, N. Suntzeff \&
J. C.  Wheeler, Astrophys. J. Lett., {\bf 509}, L117-L120 (1998). (I, A)

\item 
``The Spectra of the Supernovae in IC 4182 and in NGC 1003," R. Minkowski,
Astrophys. J., {\bf 89}, 156-217 (1939).(E, I)

\item
``Spectra of Supernovae," R. Minkowski, Pub. Astronom. Soc. Pac.,
\textbf{53},
224-225 (1941). (E, I)

\item
``Is the Compact Source in the Center of Cassiopeia A Pulsed?"
S. S. Murray, S. M. Ransom, M. Juda, U. Hwang \& S. S. Holt,
Astrophys. J., {\bf 566}, 1039-1044 (2002). (I, A)

\item
``Type I Supernovae Come From Short-lived Stars," A. Oemler, Jr. \&
B. M. Tinsley, Astronom. J., {\bf 84}, 985-992 (1979). (E. I)

\item
``Multifrequency Observations of Recent Supernovae," N. Panagia, Lect.
Notes Phys., {\bf 224}, 14-33 (1985). (I, A)

\item
``Measurements of Omega and Lambda from 42 High-Redshift Supernovae,"
S. Perlmutter, G. Aldering, G. Goldhaber, R. A. Knop,
P. Nugent, P. G. Castro, S. Deustua, S. Fabbro, A. Goobar,
D. E. Groom, I. M. Hook, A. G. Kim, M. Y. Kim, J. C. Lee,
N. J. Nunes, R. Pain, C. R. Pennypacker, R. Quimby, C. Lidman,
R. S. Ellis, M. Irwin, R. G. McMahon, P. Ruiz-Lapuente, N. Walton,
B. Schaefer, B. J. Boyle, A. V. Filippenko, T. Matheson,
A. S. Fruchter, N. Panagia, H. J. M. Newberg, \& W. J. Couch,
Astrophys. J., {\bf 517}, 565-586 (1999). (E, I, A)

\item
``The Absolute Magnitude of Type Ia Supernovae," M. M. Phillips, Astrophys.
J.  Lett., \textbf{ 413}, L105-L108 (1993). (I, A)

\item
``An Optical Spectrophotometric Atlas of Supernova 1987A in the LMC I - The
first 130 Days," M. M. Phillips, S. R. Heathcote, M. Hamuy,
\& M. Navarrete,  Astronom. J., {\bf 95}, 1087-1110  (1988). (I, A)

\item
``SN 1991T - Further Evidence of the Heterogeneous
Nature of Type Ia Supernovae," M. M. Phillips, L. A. Wells, N. B. Suntzeff,
M. Hamuy, B. Leibundgut, R. P. Kirshner, \& C. B. Foltz,  Astronom. J., {\bf
103},
1632-1637 (1992). (I, A)

\item
``The Observational Properties of Type Ib Supernovae," A. C. Porter \&
A. V. Filippenko, Astronom. J., \textbf{93}, 1372-1380 (1987). (E, I)

\item
``Light Curves, Color Curves and Expansion Velocity of Type I Supernovae as
Functions
of the Rate of Brightness Decline," Y. P. Pskovskii, Soviet Astr.,
\textbf{21},
675-682 (1977). (E, I)

\item
``GRB 970228 Revisited: Evidence for a Supernova in the Light Curve and
Late Spectral Energy Distribution of the Afterglow," D. E. Reichart,
Astrophys. J. Lett., {\bf 521}, L111-L115 (1999). (I, A)

\item
``Observational Evidence from Supernovae for an Accelerating Universe
and a Cosmological Constant," A. Riess,
A. V. Filippenko, P. Challis, A. Clocchiatti,
A. Diercks, P. M. Garnavich, R. L. Gilliland, C. J. Hogan,
S. Jha, R. P. Kirshner, B. Leibundgut, M. M. Phillips,
D. Reiss, B. P. Schmidt, R. A. Schommer, R. C. Smith,
J. Spyromilio, C. Stubbs, N. B. Suntzeff \& J. Tonry, Astronom. J., {\bf
116},
1009-1038, (1998). (E, I, A)

\item
``The Farthest Known Supernova: Support for an Accelerating Universe and a
Glimpse of the Epoch of Deceleration," A. G. Riess, P. E. Nugent,
R. L. Gilliland, B. P. Schmidt, J. Tonry, M. Dickinson,
R. I. Thompson, T. Budav‡ri, S. Casertano, A. S. Evans,
A V. Filippenko, M. Livio, D. B. Sanders, A. E. Shapley,
H. Spinrad, C. C. Steidel, D. Stern, J. Surace \&
S. Veilleux, Astrophys. J., {\bf 560}, 49-71 (2001). (I, A)

\item
``A New Subclass of Type II Supernovae?" E. M. Schlegel, Mon. Not. Roy.
Astr. Soc.,
\textbf{244}, 269-271 (1990). (I, A)

\item
``Carbon Monoxide in Supernova 1987A," J. Spyromilio, W. P. S. Meikle,
R. C. M. Learner, \& D. A. Allen, Nature, \textbf{334}, 327-329 (1988). (I,
A)

\item
``What Was Supernova 1988Z?" R. A. Stathakis, E. M. \& Sadler, Mon. Not.
Roy. Astr. Soc., {\bf 250}, 786-795 (1991). (I, A)

\item
``Catalog of 558 Pulsars," J. H. Taylor, R. N. Manchester, A. G. Lyne,
Astrophys. J. Supp., {\bf 88}, 529-568 (1993). (I, A)

\item
``Spectropolarimetry of SN 1993J in NGC 3031," S. R. Trammell, D. C. Hines
\& J. C.  Wheeler, Astrophys. J., \textbf{414}, L21-L24 (1993). (I, A)

\item
``Probing the Geometry and Circumstellar Environment of SN 1993J in M81," H.
D. Tran,
A. V. Filippenko, G. D. Schmidt, K.S. Bjorkman, B.J. Januzzi, P. S. Smith,
Pub.
Astronom. Soc. Pac.,
\textbf{109}, 489-503 (1987). (I, A)

\item 
``The Late-Time Spectrum of the Type II Supernova 1980k," A.
Uomoto, \& R. P.
Kirshner, Astrophys. J., {\bf 308}, 685-690 (1986).

\item
``The Peculiar Type II Supernova 1997D: A Case for a Very Low
$^{56}$Ni Mass," M. Turatto, P. A. Mazzali, T. R. Young, K.
Nomoto, K. Iwamoto, S. Benetti, E. Cappallaro, I. J. Danziger,
D. F. de Mello, M. M. Phillips, N. B. Suntzeff, A. Clocchiatti,
A. Piemonte, B. Leibundgut, R. Covarrubias, J. Maza \& J.
Sollerman, Astrophys. J. Lett., {\bf 498}, L129-L133
(1998). (I, A)

\item
``The Type II Supernova 1988Z in MCG+03-28-022 - Increasing Evidence of
Interaction of Supernova Ejecta with a Circumstellar Wind," M. Turatto, E.
Cappellaro,
I. J. Danziger, S. Benetti, C. Giouffes,
\& M. Della Valle,  Mon. Not. Roy. Astr. Soc., {\bf 262}, 128-140 (1993).
(I, A)

\item
``Rediscussion of Extragalactic Supernova Rates Derived from Evan's
1980-1988
Observations," S. van den Bergh,
\& R. D. McClure, Astrophys. J., {\bf 425}, 205-209 (1994). (I, A)

\item
``A Catalog of Recent Supernova," S. van den Bergh, Astrophys. J. Supp.,
{\bf
92}, 219-227 (1994). (I, A)

\item 
``Association of Supernova with Recent Star Formation Regions in Late Type
Galaxies," S. D. Van Dyk, Astronom. J., {\bf 103}, 1788-1803 (1992). (I, A)

\item
``Transient Optical Emission from the Error Box of the Gamma-Ray Burst
of 28 February 1997," J. van Paradijs, P. J. Groot,
T. Galama, C. Kouveliotou, R. G. Strom, J. Telting, R. G. M. Rutten,
G. J. Fishman, C. A. Meegan, M. Pettini, N. Tanvir, J. Bloom,
H. Pedersen, H. U. Nordgaard-Nielsen, M. Linden-Vornle, J. Melnick,
G. vanÊderÊSteene, M. Bremer, R. Naber, J. Heise, J. inÊ'tÊZand,
E. Costa, M. Feroci, L. Piro, F. Frontera, G. Zavattini, L. Nicastro,
E. Palazzi, K. Bennet, L. Hanlon \& A. Parmar, Nature, {\bf 386}, 686-689
(1997). (I, A)

\item
``Observations of the Nebulosities Near SN 1987A," E. J. Wampler, L. Wang,
D. Baade, K. Banse, S. D'Odorico, C. Gouiffes \& M. Tarenghi,
Astrophys. J. Lett., {\bf 362}, L13-L16 (1990). (I, A)

\item
``Bipolar Supernova Explosions," L. Wang, D. A. Howell, P. H\"oflich,
\& J. C. Wheeler, Astrophy. J., {\bf 550}, 1030-1035 (2001). (I, A)

\item 
``Supernovae 1983I and 1983V - Evidence for Abundance
Variations in Type Ib Supernovae," J. C. Wheeler, R. P.
Harkness, E. S. Barker, A. L. Cochran,
\& D. Wills, Astrophys. J. Lett., {\bf 313}, L69-L73 (1987). (I, A)

\item
``The Peculiar Type I Supernova in NGC 991," J. C. Wheeler, and R.
Levreault,
Astrophys. J. Lett., \textbf{294}, L17-L20 (1985). (I, A)

\item
``SN 1994I in M51 and the Nature of Type IBC Sopernovae," J. C. Wheeler, R.
P. Harkness, A. Clocchiatti, S. Benetti, D. Depoy, \& J. Elias, Astrophys.
J.  Lett., \textbf{436}, L135-L138 (1994). (I, A)

\subsection*{ IV.2 Theory}

\item
``A Possible Model of Supernovae: Detonation of 12C," W. D. Arnett,
Astrophys. \& Sp. Sci., {\bf 5}, 180-212 (1969). (I, A)

\item
``Neutrino Trapping During Gravitational Collapse of Stars," W. D. Arnett,
Astrophys. J., \textbf{218}, 815-833 (1977). (I, A)

\item
``Supernova Theory and Supernova 1987A,"  W. D. Arnett, Astrophys. J.,
\textbf{319},
136-142 (1987). (I, A)

\item
``The Cosmic Triangle: Revealing the State of the Universe,"
N. A. Bahcall, J. P. Ostriker, S. Perlmutter \& P. J. Steinhardt,
Science, {\bf 284}, 1481-1488 (1999). (E, I)

\item
``Dynamics of Supernova Explosions Resulting from Pair Formation," Z.
Barkat,
G. Rakavy, and N. Sack, Phys. Rev. Lett., \textbf{18}, 379-381 (1967). (I,
A)

\item
``Collapse of 9 Solar Mass Stars," E. Baron, J. Cooperstein, \& S. Kahana,
Astrophys. J., \textbf{320}, 300-303 (1987). (I, A)

\item
``Supernova Shock. VIII," H.A. Bethe, Astrophys. J., \textbf{490}, 765
(1997). (I, A)

\item
``Equation of State in the Gravitational Collapse of Stars," H. A. Bethe,
G. E. Brown, J. Applegate. \& J. M. Lattimer, Nucl. Phys.  A.,
{\bf 324}, 487-533 (1979). (I, A)

\item
``Type I Supernovae," D. Branch \& B. Patchett, Mon. Not. Roy. Astr. Soc.,
161, 71-83 (1973). (I, A)

\item
``Stellar Core Collapse - Numerical Model and Infall Epoch," S. W. Bruenn,
Astrophys. J. Supp., \textbf{58}, 771-841 (1985). (A)

\item
``On the Nature of Core-Collapse Supernova Explosions," A. Burrows, J.
Hayes, \&
B. A. Fryxell, Astrophys. J., \textbf{450}, 830-850 (1995). (I, A)

\item
``Neutron Star Accretion in a Supernova," R. A. Chevalier, Astrophys. J.,
{\bf 346}, 847-859 (1989). (I, A)

\item
``Indications of a Superwind in the Lightcurves of Type-II Supernovae,"
N. N. Chugai, Sov. Ast., {\bf 36}, 63-69 (1992). (I, A)

\item
``Early Supernova Luminosity," S. A. Colgate, \& C. McKee. Astrophys. J.,
\textbf{157}, 623-644 (1969). (I, A)

\item
``The Hydrodynamic Behavior of Supernovae Explosions," S. A. Colgate, \& R.
H. White,
Astrophys. J., \textbf{143}, 626-681 (1966). (I, A)

\item
``Explosions in Wolf-Rayet Stars and Type Ib Supernovae: I- Light Curves,"
L.
M. Ensman, and S. E. Woosley, Astrophys. J., \textbf{333}, 754-776 (1988).
(I, A)

\item
``Lepton-Driven Convection in Supernovae," R. I. Epstein, Mon. Not. Roy.
Astr.
Soc., \textbf{188}, 305-325 (1979). (I, A)

\item
``A Theoretical Model for Type II Supernovae," S. W. Falk \& D. W. Arnett,
Astrophys. J. Lett., {\bf 180}, L65-L68 (1973). (I, A)

\item 
``The Freeze-Out Phase of SN 1987A - Implications for the Light Curve," C.
Fransson
and C. Kozma, Astrophys. J., {\bf 408}, L25-L28 (1993). (I, A)

\item
``Core-Collapse Simulations of Rotating Stars," C. L. Fryer \& A. Heger,
Astrophys. J., {\bf 541}, 1033-1050 (2000). (I, A)

\item
``Homologously Collapsing Stellar Cores," P. Goldreich \& S. V. Weber,
Astrophys. J., {\bf 238}, 991-997 (1980). (A)

\item
``The Cooling Wave in Supernova Shells," E. K. Grasberg \& D. K. Nadezhin,
Astrophys. \& Sp. Sci., {\bf 44}, 409-428 (1976). (E, I, A)

\item
``A New Model for Progenitor Systems of Type Ia Supernovae,"
I. Hachisu, M. Kato \& K. Nomoto, Astrophys. J. Lett., {\bf 470},
L97-L100 (1996). (I, A)

\item
``The Optical Radiation of Supernovae," R. P. Harkness, in {\bf Radiation
Hydrodynamics in Stars and  Compact Objects}, ed. D. Mihalas and K.-H.A.
Winkler 
(Berlin: Springer-Verlag 1986), pp. 166-181. (I, A)

\item
``Postcollapse Hydrodynamics of SN 1987A - Two Dimensional Simulations of
the Early
Evolution," M. Herant, W. Benz,
\& S. A. Colgate, Astrophys. J., \textbf{395}, 642-653 (1992). (I, A)

\item
``Inside the Supernova: A Powerful Convective Engine," M. Herant,
W. Benz, W. R. Hix, C. L. Fryer \& S. A.  Colgate, Astrophys. J., {\bf 435},
339-361 (1994). (I, A)

\item
``Supernova Explosions of Massive Stars - The Mass Range 8 to 10 Solar
Masses,"
W. Hillebrandt, A. G. Wolff \& K. Nomoto, Astron. \& Astropys., {\bf 133},
175-184 (1984). (I, A)

\item
``Remarks on the Polarization Observed in SN 1993J," P. H\"oflich,
Astrophys. J., 
\textbf{440}, 821-829 (1995). (A)

\item
``Explosion Models for Type Ia Supernovae: A Comparison with Observed
Light Curves, Distances, H$_0$, and q$_0$," P. H\"oflich \& A. Khokhlov,
Astrophys. J., {\bf 457}, 500-528 (1996). (I, A)

\item
``Delayed Detonation Models for Normal and Subluminous Type Ia Supernovae:
Absolute Brightness, Light Curves, and Molecule Formation," P. H\"oflich,
A. M. Khokhlov \& J. C. Wheeler, Astrophys. J., {\bf 444}, 831-847 (1995).
(I, A)

\item
``Maximum Brightness and Postmaximum Decline of Light Curves of Type Ia
Supernovae: A Comparison of Theory and Observations," P. H\"oflich,
A. Khokhlov, J. C. Wheeler, M. M. Phillips, N. B. Suntzeff \& M.  Hamuy,
Astrophys. J. Lett., {\bf 472}, L81-L84 (1996). (I, A)

\item
``On the Thermonuclear Runaway in Type Ia Supernovae: How to run away?"
P. H\"oflich \& J. Stein, Astrophys. J., {\bf 568}, 779-790 (2002). (I, A)

\item
``Extreme Type IIp Supernovae as Yardsticks for Cosmology," P. H\"oflich,
O. Straniero, M. Limongi, I. Domingue \& A. Chieffi,
Rev. Mex. de Astronom'a y Astrof'sica, {\bf 10}, 157-162 (2001). (I, A)

\item
``Hard X-Rays and Gamma Rays from Type Ia Supernovae," P. H\"oflich,
J. C. Wheeler \& A. Khokhlov, Astrophys J., \textbf{492}, 228-245 (1998).
(I, A)

\item
``Nucleosynthesis in Supernovae," F. Hoyle \& W. A. Fowler, Astrophys. J.,
\textbf{132}, 565-590 (1960). (E, I)

\item
``Nucleosynthesis in Chandrasekhar Mass Models for Type Ia Supernovae
and Constraints on Progenitor Systems and Burning-Front Propagation,"
K. Iwamoto, F. Brachwitz, K. Nomoto, N. Kishimoto, H. Umeda,
W.R. Hix \& F.-K. Thielemann, Astrophys. J. Supp., \textbf{125}, 439-462
(1999). (I, A) 

\item
``Conditions for Shock Revival by Neutrino Heating in Core-Collapse
Supernovae," H.-Th. Janka, Astr. \& Astrophys., {\bf 368}, 527-560 (2001).
(A)

\item
``The First Second of Type II Supernova: Convection, Accretion, and Shock
Propagation,"  H.-T. Janka,
\& E. M\"uller, Astrophys. J. Lett., \textbf{448}, L109-L114 (1995). (A)

\item
``Neutrino Heating, Convection, and the Mechannism of Type-II Supernova
Explosions," H.-T. Janka,
\& E. M\"uller, Astr. \& Astrophys., \textbf{306}, 167-198 (1996). (A)

\item
``Analysis of SN 1987A Polarimetry," D. J. Jeffery, Astrophys. J.,
\textbf{375},
264-287 (1991). (A)

\item
``Delayed Detonation Model for Type Ia Supernovae," A. M. Khokhlov,
Astr. \& Astrophys., \textbf{245}, 114-128 (1991). (I, A)

\item
``Propogation of Turbulent Flames in Supernovae," A. M. Khokhlov, Astrophys.
J.,
\textbf{449}, 695-713 (1995). (I, A)

\item
``Three-Dimensional Modeling of the Deflagration Stage of a
Type Ia Supernova Explosion," A. M. Khokhlov, Astrophys. J., in press
(2002), (astro-ph/0008463).  (A)

\item
``Jet-Induced Explosions of Core Collapse Supernovae," A. M. Khokhlov,
P.A. H\"oflich, E. S. Oran, J. C. Wheeler, L. Wang \& A. Yu. Chtchelkanova,
Astrophys. J., \textbf{524}, L107-L110 (1999). (I, A)

\item
``Deflagration-to-Detonation Transition in Thermonuclear Supernovae," A. M.
Khokhlov,
E. S. Oran,
\& J. C. Wheeler, Astrophys. J., \textbf{478}, 678-688 (1997). (I, A)

\item
``The Evolution of Main Sequence Star + White Dwarf Binary
Systems Towards Type Ia Supernovae," N. Langer, A. Deutschmann,
S. Wellstein \& P. H\"oflich, Astr. \& Astrophys.,
{\bf 362}, 1046-1064 (2000). (I, A)

\item
``A Generalized Equation State for Hot, Dense Matter," J. M. Lattimer, \& F.
D.
Swesty, Nucl. Phys., \textbf{535}, 331-376 (1991). (I, A)

\item
``A Numerical Example of the Collapse of a Rotating Magnetized Star," J. M.
LeBlanc,
\& J. R. Wilson, Astrophys. J., \textbf{161}, 541-552 (1970). (I, A)

\item
``SN 1984A and Delayed-Detonation Models of Type Ia Supernovae,"
E. J. Lentz, E. Baron, D. Branch \& P. H. Hauschildt, Astrophys. J.,
{\bf 547}, 402-405 (2001). (A)

\item
``Supernovae\hspace{.5em}from\hspace{.5em}Collapse\hspace{.5em}of
\hspace{.5em}Oxygen\hspace{.5em}-\hspace{.5em}Magnesium\hspace{.5em}
-\hspace{.5em}Neon\hspace{.5em} Cores," R. W. Mayle, \& J. R. Wilson, 
Astrophys. J., \textbf{334}, 909-926 (1988). (A)

\item
``Type II Supernova and Boltzman Neutrino Transport - The Final Phase," A.
Mezzacappa,
\& S. W. Bruenn, Astrophys. J., \textbf{405}, 637-668 (1993). (A)
 
\item
``An Investigation of Neutrino-Driven Convection and the Core Collapse
Supernova Mechanism Using Multigroup Neutrino Transport," A. Mezzacappa, A.
C. Calder,
S. A. Bruenn, J. M. Blondin, M. W. Guidry, M. R. Strayer, \& A. S. Umar,
Astrophys. J., \textbf{495}, 911-926 (1998). (A)

\item
``Simulation of the Spherically Symmetric Stellar Core Collapse, Bounce,
and Postbounce Evolution of a Star of 13 Solar Masses with Boltzmann
Neutrino Transport, and Its Implications for the Supernova Mechanism,"
A. Mezzacappa, M. Liebendšrfer, O. E. Messer, R. W. Hix, F.-K. Thielemann,
\& S. W. Bruenn, Phys. Rev. Lett., {\bf 86}, 1935-1938 (2001). (A)

\item
``High-Redshift Supernovae and the Metal-poor Halo Stars:
Signatures of the First Generation of Galaxies," J. Miralda-Escud\'e
\& M. J. Rees, Astrophys. J., {\bf 478}, L57-L61 (1997). (I, A)

\item
``Gravitational Waves from the Collapse of Rotating Stellar Cores," R.
M\"onchmeyer,
G. Sch\"afer, E. M\"uller,
\& R. E. Kates, Astr. \& Astrophys., \textbf{246}, 417-440 (1991). (A)

\item
``The Effect of Neutrino Transport on the Collapse of Iron Stellar Cores,"
E. S. Myra, S.A. Bludman, Y. Hoffman, I. Lichtenstadt, N. Sack, Astrophys.
J., \textbf{318}, 744-759 (1987). (A)

\item
``The Thermonuclear Explosion of Chandrasekhar Mass White Dwarfs,"
J. C. Niemeyer \& S. E. Woosley, Astrophys. J., {\bf 475},
740-753 (1997). (E, I, A)

\item
``Evolution of 8-10 Solar Mass Stars Toward Electron Capture Supernovae.
I - Formation of Electron-Degenerate O + Ne + Mg Cores," K. Nomoto,
Astrophys. J., {\bf 277}, 791-805 (1984). (I, A)

\item
``Fundamental Physical Parameters of Collimated Gamma-Ray Burst Afterglows,"
A. Panaitescu \& P. Kumar, Astrophys. J. Lett., {\bf 560}, L49-L53 (2001).
(I, A)

\item
``Possible Thermonuclear Activities in Natural Terrestrial Minerals," T.
Pankey,
  Thesis (Ph.D.) Howard University Diss. Abs. Int'l., \textbf{23} (04), 1395
(1962). 

\item
``Anomalous Beta Decay in Type-I Supernovae," T. Pankey Jr., Pub. Astr. Soc.
Pac.,
\textbf{92}, 606-608 (1980).  (E. I)

\item
``The Physics of Type Ia Supernova Light Curves. II. Opacity and
Diffusion," P. A. Pinto \& R. G. Eastman, Astrophys. J., {\bf
530}, 757-776 (2000). (I, A)

\item
``The Progenitor of SN 1987A," P. Podsiadlowski, Pub. Astr. Soc. Pac.,
{\bf 104}, 717-729 (1992). (I, A)

\item
``Identification of the Absorption Spectrum of the Type I Supernova," Yu. P.
Pskovskii,
Soviet Astronomy, {\bf 12}, 750-756 (1969). (I, A)

\item
``Refined Numerical Models for Mulltidimensional Type Ia Supernova
Simulations," M. Reinecke, W. Hillebrandt \& J. C. Niemeyer,
Astr. \& Astrophys., \textbf{386}, 936-943 (2002). (A)

\item
``Effect of Anisotropic Neutrino Radiation on Supernova Explosion Energy,"
T. M. Shimizu, T. Ebisuzaki, K. Sato \& S. Yamada, Astrophys. J., {\bf 552},
756-781 (2001). (I, A)

\item
``Models for Type I Supernovae- Partially Incinerated White Dwarfs," P. G.
Sutherland,
\&  J. C. Wheeler, Astrophys. J., \textbf{280}, 282-297 (1984). (I, A)

\item
``Core-Collapse Supernovae and Their Ejecta," F.-K. Thielemann, K. Nomoto
\& M. Hashimoto, Astrophys. J., {\bf 460}, 408-436 (1996). (I, A)

\item
``Explosive Nucleosynthesis in Carbon Deflagration Models of Type I
Supernovae,"
F.-K. Thielemann, K. Nomoto \& K. Yokoi, Astr. \& Astrophys., {\bf 158},
17-33
(1986). (I, A)

\item
``Galactic\hspace{.5em}Chemical\hspace{.5em}Evolution:\hspace{.5em}Hydrogen\
hspace{.5em}Through\hspace{1mm}Zinc,"
F. X. Timmes, S. E. Woosley
\& T. A. Weaver, Astrophys. J. Supp., {\bf 98}, 617-658 (1995).
(I, A)

\item
``Explosion Diagnostics of Type Ia Supernovae from Early Infrared Spectra,"
J. C. Wheeler, P. H\"oflich, R. P. Harkness \& J. Spyromilio,
Astrophys. J. Lett., {\bf 496}, 908-914 (1998). (I, A)

\item
``Binaries and Supernovae of Type I," J. Whelan \& I. Iben, Jr.,
Astrophys. J., {\bf 186}, 1007-1014 (1973). (I, A)

\item
``Perspectives on Particle Physics and Cosmology," F. Wilczek, Physica
Scripta,
{\bf 36}, 281-296 (1991).  (E, I, A)

\item
``Gamma-ray Bursts from Stellar Mass Accretion Disks Around Black Holes,"
S. E. Woosley, Astrophys. J., {\bf 405}, 273-277 (1993). (I, A)

\item
``Models for Type I Supernova. I - Detonations in White Dwarfs," S. E.
Woosley,
R. E. Taam, and T. A. Weaver, Astrophys. J., \textbf{301}, 601-623 (1986).
(I, A) 

\item
``SN 1987A - After the Peak," S. E. Woosley, Astrophys. J., \textbf{330},
218-253
(1988). (I, A)

\end{enumerate}

\end{document}